\documentclass[
nofootinbib,
amsmath,amssymb,
aps,
prd,
preprint,
]{revtex4-1}

\pdfoutput=1
\usepackage[dvipsnames]{xcolor}
\usepackage{graphicx}
\usepackage{dcolumn}
\usepackage{bm}
\usepackage{natbib}
\usepackage{tikz}

\usepackage{hyperref}
\usepackage{braket}

\newcommand{\sxx}{\braket{S^{(1)}_xS^{(2)}_x}}
\newcommand{\szz}{\braket{S^{(1)}_zS^{(2)}_z}}

\begin{document}

\title{Statistical Bounds on CMB Bell Violation}

\newcommand{\SECONDAFF}{\affiliation{Department of Physics,
	University at Buffalo, SUNY
	Buffalo,
	NY 14260
	USA}}

\author{Michael J. P. Morse}
\email{mjmorse3@buffalo.edu}
\SECONDAFF

\date{\today}

\begin{abstract}
Inspired the low $\ell$ cosmic variance we present a study on the statistical variance expected in observing Bell's inequality violation by primordial quantum states violation. We consider the statistical variance inherent to three sets of pseudo-spin operators used to construct Bell's inequality. We find that for a highly squeezed state, such as the relevant CMB states, the statistical variation will vanish and does not contribute to theoretical uncertainties.  
\end{abstract}

\maketitle

\section{Cosmic Bell's Introduction}

The origin of large scale structure and the anisotropies on the Cosmic Microwave Background (CMB) are explained in standard inflationary cosmology as the result of quantum fluctuations of the inflation field \cite{Starobinsky:1980te,PhysRevD.23.347,10.1093/mnras/195.3.467,Albrecht:1982wi,Linde:1981mu,Linde:1983gd,Ade:2013zuv,Planck:2013jfk,Ade:2013ydc,Ade:2015xua,Ade:2015lrj,Ade:2015ava}. The quantum mechanical perturbations are imprinted on the CMB as classical temperature anisotropies. Since the classical temperature anisotropies started out as quantum mechanical in nature, it has been suggested that quantum mechanical phenomena may be observable. In particular, violations of Bell's inequalities \cite{PhysicsPhysiqueFizika.1.195,Martin:2016tbd,Martin:2017zxs}. 

Bell's inequalities demonstrate that the fundamental laws of nature can not be described by a theory which is both local and deterministic \cite{Hensen:2015ccp,PhysRevLett.81.5039,PhysRevLett.49.1804,PhysRevLett.49.91,PhysRev.47.777}. Although Bell's inequalities were originally applied to entangled particles in position space, typically the Clauser-Horne-Shimony-Holt (CHSH) formalism is considered. The CHSH formalism has the property that the eigenvalues of the system are discrete, simplifying the inequalities \cite{PhysRev.47.777,PhysRevLett.23.880}. 

Recently methods have been developed to transform continuous systems into discrete ones in order to apply the CHSH inequalities. In this work we consider three sets of these \textit{pseudo}-spin operators, Banaszek-Wodkiewicz (BW) operators, Gour-Khanna-Mann-Revzen (GKMR) operators, and Larsson operators \cite{PhysRevLett.82.2009,PhysRevLett.88.040406,Revzen2006-REVTWF,PhysRevA.71.022103,GOUR2004415,PhysRevA.70.022102}. The construction of discrete pseudo-spin operators from continuous operators opens the door to testing the CHSH inequality on more systems. In our case, this development allows one to construct pseudo-spin operators out of the Fourier amplitude of the inflationary quantum modes frozen on the CMB. 

At the end of inflation the inflaton fluctuations are in a two mode squeezed state, testable for CHSH violation \cite{PhysRevD.42.3413,PhysRevD.46.1440,Polarski_1996,MUKHANOV1992203,Mukhanov:1981xt,lemoine2007inflationary,amelino2005planck}. One may wonder if an experiment could be constructed that probes for Bell's inequality violation on the CMB.  This question is addressed in \cite{Martin:2016tbd,Martin:2017zxs}, where the authors construct the relevant CHSH operators and calculate the expectation value. However, since quantum mechanical observables are statistical in nature, just because the mean value violates Bell's inequality does not necessarily imply that the observed realization will. If we had an infinite sized sky on which to preform the experiment many times we would expect the mean of the observed values to approach the theoretical mean. However, the CMB has a fixed number of observable modes at any frequency, limiting the statistics. If violation of Bell's inequality manifests for modes that primarily contribute to low multipole moments, observation of the violation may be \textit{cosmic variance limited}. It is therefore prudent to check if violation would still be visible within a few $\sigma$ around the theoretical mean. 

 CHSH inequality violation has an upper bound, the Cirel'son bound \cite{Cirelson:1980ry}, at $2 \sqrt{2}$ and a lower ``classical bound" at $2$ leading to a narrow parameter space of observable violation. This is not an issue in the typical laboratory application of the inequality since the experiment can be ran many times so that the variance will drop off proportional to the number of trials. However, on the CMB we are limited by the number of patches of the sky which can be observed, potentially limiting the statistics. In this paper we will calculate the statistical variance associated with the CHSH operator and show that in the highly squeezed limit the statistical variance vanishes irrespective of the sample size. 

The paper is structured as follows. Section \ref{sec:Bell} will review the work done in \cite{Martin:2016tbd,Martin:2017zxs} to set up the CHSH operator and quantum state. In Section \ref{sec:variance} we will derive the variance associated with the two point spin operators as well as an expression for the uncertainty on the expectation value of the Bell operator. In Section \ref{sec:pseudo-spin} we set up the pseudo-spin operators and apply the results of Section \ref{sec:variance} to calculate the associated variance. Section \ref{sec:conclusion} will present our conclusion.  
\section{CHSH Operator}
\label{sec:Bell}
In this section we will review the Clauser-Horne-Shimony-Holt formulation of Bell's inequalities. This section follows from \cite{PhysRevLett.23.880,Martin:2016tbd,Martin:2017zxs}.

Consider a two particle state whose Hilbert space can be written as the direct product of one particle states,
\begin{align}
    \mathcal{H} = \mathcal{H}_1 \otimes \mathcal{H}_2.
\end{align}
 Suppose one measures the spin of particle one with respect to some direction in the $(x,z)$\footnote{In principle we could include the azimuthal angle $\phi$, however we can set $\phi = 0$ WOLOG.} plane, $\mathbf{n} = (\sin(\theta_1), \cos(\theta_1))$, and particle two with respect to some direction $\mathbf{m}=(\sin(\theta_2), \cos(\theta_2))$, then repeats the measurements along directions $\mathbf{n'} = (\sin(\theta'_1), \cos(\theta'_1))$, $\mathbf{m'} = (\sin(\theta'_2), \cos(\theta'_2))$. With these measurements the CHSH operator is
\begin{align}
\label{eq:B}
\mathcal{B} \equiv \left[ \mathbf{n} \cdot \hat{\mathbf{S}}_1 \right] \otimes  \left[\mathbf{m} \cdot \hat{\mathbf{S}}_2  \right] + 
\left[ \mathbf{n'} \cdot \hat{\mathbf{S}}_1 \right] \otimes  \left[\mathbf{m} \cdot \hat{\mathbf{S}}_2  \right]  +
\left[ \mathbf{n} \cdot \hat{\mathbf{S}}_1 \right] \otimes  \left[\mathbf{m'} \cdot \hat{\mathbf{S}}_2  \right] -
\left[ \mathbf{n'} \cdot \hat{\mathbf{S}}_1 \right] \otimes  \left[\mathbf{m'} \cdot \hat{\mathbf{S}}_2  \right], 
\end{align}
 where $\hat{\mathbf{S}}$ is a vector of spin operators along $x$ and $z$, 
\begin{align}
    \hat{\mathbf{S}} = \left(\mathbf{\hat{S}_x},\mathbf{\hat{S}_z} \right).
\end{align}If the angles $\theta_{1},\theta'_{1},\theta_{2},\theta'_{2}$ are optimally chosen to maximize the expectation value of Equation (\ref{eq:B}) the expression becomes, 
\begin{align}
\mathcal{B} &= \frac{2}{\braket{\mathcal{B}}}\left\lbrace \left( \hat{S}^{(1)}_z + \hat{S}^{(1)}_x \right) 
\otimes \left(\sxx \hat{S}^{(2)}_x + \szz \hat{S}^{(2)}_z  \right) -\right. \nonumber
\\
&\left. 
\left( \hat{S}^{(1)}_z - \hat{S}^{(1)}_x \right) \otimes \left(\sxx \hat{S}^{(2)}_x - \szz \hat{S}^{(2)}_z   \right) \right \rbrace.
\label{eq:CHSH}
\end{align}
 The expectation value of Equation (\ref{eq:CHSH}) has a simple expression, 
\begin{align}
    \braket{\mathcal{B}} &= 2 \sqrt{\sxx^2 + \szz^2}. \label{eq:<CHSH>}
\end{align}
One notices that since $\mathbf{\hat{S}_{x,y}}$ are spin operators,
\begin{align}
    \sxx^2,\szz^2 \le 1. 
\end{align}
The maximal violation, given by the Cirel'son bound, is recovered when the inequalities are saturated
\begin{align}
    \braket{\mathcal{B}} = 2 \sqrt{2}. 
\end{align}

In this paper the two-mode squeezed state we consider is
\begin{align}
\ket{\Psi_{2,sq}} = \frac{1}{\cosh(r_k)}\sum_{n=0}^{\infty}(-1)^n\mathrm{e}^{2in\varphi_k}\tanh(r_k)^n \ket{n_k,-n_k},
\label{eq:squeeze_state}
\end{align}
since this is the state in which cosmological fluctuations are in at the end of inflation \cite{Leach:2001zf,Henderson_2001,Martin:2016tbd}. $r_k,\varphi_k$ are the squeeze parameter and squeeze angle respectively which give a measure of how entangled the state is. On the super-Horizon scales $\mathbf{k} \ll aH$ we are interested in, $r_k$ is on the order of the number of e-folds spend outside the Hubble radius, $r_k \sim 50$, and $\varphi_k = -\pi/2$. The large value of $r$ indicates that the modes under consideration are highly squeezed.

\section{Variance}
\label{sec:variance}
We consider the variance of the Bell inequality, in CHSH form, defined as,
\begin{align}
\sigma^2_\mathcal{B} = \braket{\mathcal{B}^2} - \braket{\mathcal{B}}^2.
\label{eq:var}
\end{align}
There may be a more optimal unbiased operator to use for the variance, however for our application the naive operator will suffice to put an upper bound on the expected variance, a more optimal choice would only act to strengthen the bound. Equation (\ref{eq:var}) is recognizable as the second central moment of the CHSH operator, $\mathcal{B}$. It gives us the expected spread on a single measure of the mean value, $\braket{\mathcal{B}}$. However, in practice we will have more than a single independent measurement. If we repeat the measurement $N$ times, the repeated measurement leads to a further reduction of the variance proportional to $\sqrt{N}$,
\begin{align}
\Delta_{\mathcal{B}} = \frac{1}{\sqrt{N}}\sigma_\mathcal{B}.
\label{eq:samp_var}
\end{align}
 In other words, $\Delta_{\mathcal{B}}$ takes into account the sampling variance, commonly referred to as the variation of the mean. Since we are considering entangled momentum states of perturbations formed during inflation there exits a finite number of samples, at a given frequency, that are observable on any finite sized patch of the sky. Therefore, for a given k-mode, in practice the sampling variance will be smaller than the statistical variance associated with sampling a distribution. 
 
 As we will see, for highly squeezed states in almost all cases it is sufficient to use the statistical variance, (\ref{eq:var}), over the sampling variance, (\ref{eq:samp_var}), to calculate the variance. This is due to our result that as $r \rightarrow \infty$ the variance approaches zero for the optimally chosen angles, leading to the same qualitative result. 

In order to find an expression for Equation (\ref{eq:var}) we need expressions for $\mathcal{B}$, and $\mathcal{BB}$. In the ideal case, $\mathcal{B}$ has already been obtained by the authors of \cite{Martin:2016tbd}, therefore we only need to determine the form of $\mathcal{B}^2$, and the second moment $\braket{\mathcal{B}^2}$. For optimally chosen angles,
\begin{align}
\mathcal{BB} &= 4 \left(\mathbb{I}\otimes\mathbb{I} \right) - 32 \frac{\sxx\szz}{\braket{\mathcal{B}}^2} \left(\hat{S}^{(1)}_y  \otimes \hat{S}^{(2)}_y  \right),
\label{eq:CHSH^2} \\
\braket{\mathcal{BB}} &=  4  - 32 \frac{\sxx\szz\braket{\hat{S}^{(1)}_y\hat{S}^{(2)}_y}}{\braket{\mathcal{B}}^2},
\label{eq:<CHSH^2>}
\end{align} 
where the operators all have their parameter lists omitted for ease of reading. While the operator $\mathcal{B}$ depends only on the two point function in the $x$ and $y$ directions, the operator  $\mathcal{BB}$ depends on all three two point correlation functions. The introduction of the $y$-direction two point correlation function is a result of the  $\textit{SU}(2)$ algebra governing the operators. 

Substituting these results into Equation (\ref{eq:var_CHSH}), the statistical variance is,
\begin{align}
\sigma^2_\mathcal{B} = 4  - \left( 32 \frac{\sxx\szz\braket{\hat{S}^{(1)}_y\hat{S}^{(2)}_y}}{\braket{\mathcal{B}}^2} +\braket{\mathcal{B}}^2\right). \label{eq:var_CHSH}
\end{align}

The next section is dedicated to evaluating three different sets of pseudo-spin operators and their two point functions in order to put a bound on the statistical variance.
 
\section{Pseudo-Spin Operators}
\label{sec:pseudo-spin}
The pseudo-spin operators we consider in this paper are: Banaszek-Wodkiewicz (BW) operators, Gour-Khanna-Mann-Revzen (GKMR) operators, and Larsson operators \cite{PhysRevLett.82.2009,PhysRevLett.88.040406,Revzen2006-REVTWF,PhysRevA.71.022103,GOUR2004415,PhysRevA.70.022102,Martin:2016tbd,Martin:2017zxs}. While the first two (BW, GKMR) give analytic expressions for the two point functions, the Larsson two point functions need to be numerically evaluated. The code used to numerically evaluate the Larsson operators as well as generate all the plots is publicly available on github \cite{michael_j_p_morse_2020_3688751}. 

\subsection{Banaszek-Wodkiewicz (BW) operators}  
Pseudo-spin operators give a way to produce from operators with a continuous spectrum of eigenvalues, discrete eigenvalues of $\pm 1$, the fictitious \textit{spin} value. For any $k$ mode the BW operators which will map the momentum states into pseudo-spin are \cite{PhysRevLett.82.2009,Martin:2017zxs}, 
\begin{align}
&\hat{s}_x(\mathbf{k}) = \sum_{n=0}^\infty \left( \ket{2 n_\mathbf{k} +1 }\bra{2n_\mathbf{k}} + \ket{2 n_\mathbf{k}  }\bra{2n_\mathbf{k}+1}\right), 
\\
&\hat{s}_y(\mathbf{k}) = i\sum_{n=0}^\infty \left( \ket{2 n_\mathbf{k}  }\bra{2n_\mathbf{k}+1} - \ket{2 n_\mathbf{k} +1 }\bra{2n_\mathbf{k}}\right),
\\
&\hat{s}_z(\mathbf{k}) = \sum_{n=0}^\infty \left( \ket{2 n_\mathbf{k} +1 }\bra{2n_\mathbf{k}+1} + \ket{2 n_\mathbf{k}  }\bra{2n_\mathbf{k}}\right).
\end{align}  
 Similar expressions exist for $-\mathbf{k}$ modes.  Where $\ket{2 n_\mathbf{k}}$ are the eigenvectors of the number operators which appear in the bipartite squeezed state, Equation (\ref{eq:squeeze_state}). It is not too difficult to shown that these operators obey the \textit{SU}(2) algebra. 

To evaluate (\ref{eq:var_CHSH}) we must know the expectation value of the three pseudo-spin operators in the bipartite squeezed state (\ref{eq:squeeze_state}). The $\hat{s}_x$ and $\hat{s}_z$ two point function have been worked out in the literature \cite{Martin:2017zxs} and the expectation value of two point function in $\hat{s}_y$ follows from the derivation of the two point function in $\hat{s}_x$. Here we will just quote all three results as:
\begin{align}
&\bra{\Psi_{Sq}}\hat{s}_x(\mathbf{-k})\otimes\hat{s}_x(\mathbf{k})\ket{\Psi_{Sq}} = \tanh(2 r_k)\cos(2 \varphi_k), 
\\
&\bra{\Psi_{Sq}}\hat{s}_y(\mathbf{-k})\otimes\hat{s}_y(\mathbf{k})\ket{\Psi_{Sq}} = -\tanh(2 r_k)\cos(2 \varphi_k),
\\
&\bra{\Psi_{Sq}}\hat{s}_z(\mathbf{-k})\otimes\hat{s}_z(\mathbf{k})\ket{\Psi_{Sq}} = 1.
\end{align}
Resulting in a remarkably simple expression for the variance,
\begin{align}
\sigma^2_\mathcal{B} =8-8 \tanh ^2(2 r) \cos ^2(2 \varphi ).
\end{align}
Since the value of $\tanh(x)$ approaches unity for $x \rightarrow \infty$ the variance will approach zero for highly squeezed states, $r \rightarrow \infty, 2 \varphi \rightarrow -\pi$.   
\begin{figure}
\includegraphics[width = 0.9\textwidth]{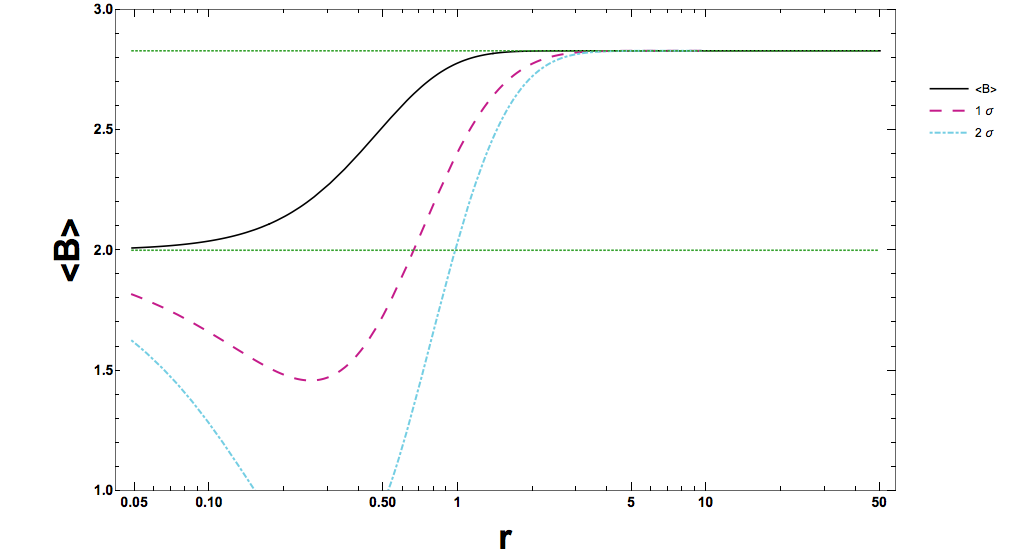}
\caption{Expectation value of CHSH violation (BW) with the one and two sigma bounds for statistical variance}
\label{fig:BW_var}
\end{figure}

As can be seen in Figure \ref{fig:BW_var} statistical variance only has a possibility of washing out the potential signal for low squeeze parameter $r$. Once $r$ is $\mathcal{O}(10)$ the statistical variance has all but disappeared and the signal is clearly visible, even for a single measurement. \textit{i.e} although the operator is statistical in nature, as the state becomes highly squeezed the expected variation in the observable goes to zero. It appears that the BW spin operators would be a good choice for probing violations on the CMB. If the violation was there we would expect to see it regardless of the number of measurements. In the next section we will consider the GKMR spin operators, for which we again have analytic forms for the spin operators, CHSH operator, and variance.

\subsection{Gour-Khanna-Mann-Revzen (GKMR) operators} 
\begin{figure}
\includegraphics[width = 0.9\textwidth]{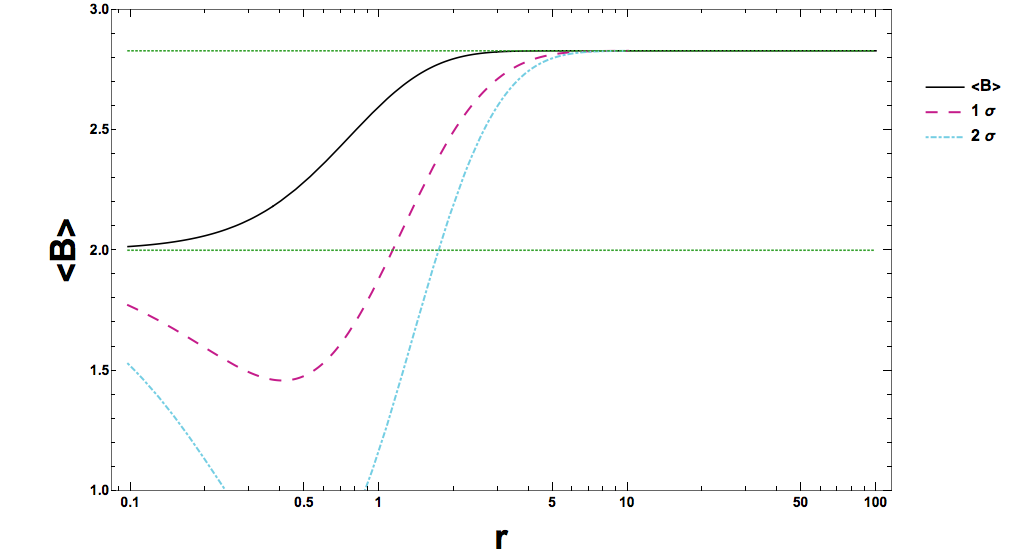}
\caption{Expectation value of CHSH violation (GKMR) with the one and two sigma bounds for statistical variance}
\label{fig:GKMR_var}
\end{figure}

Another set of analytic pseudo-spin operators are the Gour-Khanna-Mann-Revzen operators \cite{Revzen2006-REVTWF,PhysRevA.71.022103,Martin:2017zxs}. In this scheme the spin operators are
\begin{align}
 &\hat{s}_x(\mathbf{k}) =\int_0^\infty \mathrm{d}q_{\mathbf{k}} \left(\ket{q_\mathbf{k}}\bra{q_\mathbf{k}} - \ket{-q_\mathbf{k}}\bra{-q_\mathbf{k}} \right), 
\\
&\hat{s}_y(\mathbf{k}) = i \int_0^\infty \mathrm{d}q_{\mathbf{k}} \left(\ket{q_\mathbf{k}}\bra{-q_\mathbf{k}} - \ket{-q_\mathbf{k}}\bra{q_\mathbf{k}} \right),
\\
&\hat{s}_z(\mathbf{k}) = - \int_{-\infty}^\infty \mathrm{d}q_{\mathbf{k}} \left(\ket{q_\mathbf{k}}\bra{-q_\mathbf{k}}  \right),
\end{align}
where $\hat{q}_{\mathbf{k}}$ is the position operator in the Hilbert space of mode $\mathbf{k}$. The $\hat{s}_x$ and $\hat{s}_z$ two point function have been worked out in the literature \cite{Martin:2017zxs}.
\begin{align}
&\bra{\Psi_{Sq}}\hat{s}_x(\mathbf{-k})\otimes\hat{s}_x(\mathbf{k})\ket{\Psi_{Sq}} = \frac{2}{\pi} \arctan \left( \frac{2 \tanh(r_k) \cos(2 \varphi_k)}{\sqrt{1 - 2 \tanh(r_k)^2 + \tanh(r_k)^4 }}\right), 
\\
&\bra{\Psi_{Sq}}\hat{s}_y(\mathbf{-k})\otimes\hat{s}_y(\mathbf{k})\ket{\Psi_{Sq}} = -\frac{2}{\pi} \arctan \left( \frac{2 \tanh(r_k) \cos(2 \varphi_k)}{\sqrt{1 - 2 \tanh(r_k)^2 + \tanh(r_k)^4 }}\right),
\\
&\bra{\Psi_{Sq}}\hat{s}_z(\mathbf{-k})\otimes\hat{s}_z(\mathbf{k})\ket{\Psi_{Sq}} = 1.
\end{align}
With these results we can write an explicit form for the variance. The variance is not as simple as for the BW operators,
\begin{align}
&\sigma^2_\mathcal{B} = 
\tan ^{-1}\left(\frac{\sqrt{2} \left(e^{2 r}+1\right)^2 \tanh (r) \cos (2 \varphi )}{\sqrt{-\left(e^{4 r}-1\right)^2 \cos (4 \varphi )+6 e^{4 r}+e^{8 r}+1}}\right)^2 \times 
\\
&\left(\frac{32}{4 \tan ^{-1}\left(\frac{\sqrt{2} \left(e^{2 r}+1\right)^2 \tanh (r) \cos (2 \varphi )}{\sqrt{-\left(e^{4 r}-1\right)^2 \cos (4 \varphi )+6 e^{4 r}+e^{8 r}+1}}\right)^2+\pi ^2}-\frac{16}{\pi ^2}\right) \nonumber.
\label{eq:var_GKMR}
\end{align}
However, the limiting behavior is still the same as $r \rightarrow \infty,2\varphi \rightarrow -\pi$ the variance will tend to zero, as seen in Figure (\ref{fig:GKMR_var}). It can clearly be seen that, for the highly squeezed states we are interested in, there is no potential statistical limitation in observing the violation. 

The GKMR operators as well as the BW operators are a good choice since they are analytic and exhibit vanishing variance in the large squeeze limit.

The final set of operators we will consider are the Larsson operators \cite{PhysRevA.70.022102}. Unlike the previous two, BW and GKMR, the Larsson spin operators have two point functions that do not have an analytic form, and to obtain the variance the two point functions must be numerically integrated.

\subsection{Larsson operators}
The last set of pseudo-spin operators we will consider are the Larsson operators \cite{PhysRevA.70.022102}. Unlike the first two sets, these operators do not have closed form analytic representation of their two point functions. To investigate the statistical bounds on the CHSH operator we must numerically evaluate all three two point functions. The code used to do this is in the same github repository \cite{michael_j_p_morse_2020_3688751}. We include an analysis of these operators here for completeness despite the finding that, in the limit the state is highly squeezed, the power spectrum must be blue, $n_s > 1$, limiting their potential usefulness \cite{Martin:2017zxs} . 

Suppose $\ket{q_\mathbf{k}}$ is our original state which has continuous eigenvalues over the real line. The Larsson pseudo-spin operators are defined by dividing the real line into an infinite number of bins, $[nl,(n+1)l]$. Where $n$ is an integer from $-\infty$ to $\infty$ labeling the bin, and $l$ is the parameter which will define ``coarse-graining'' size. The spin operators are then defined as   
\begin{align}
 \hat{S}_x(l) &=  \hat{S}_+(l) +  \hat{S}_-(l),
 \\
 \hat{S}_y(l) &= - i [\hat{S}_+(l) -  \hat{S}_-(l)] ,
 \\
 \hat{S}_z(l) &= \sum_{n=-\infty}^{\infty} (-1)^n \int_{nl}^{(n+1)l}\mathrm{d}\mathbf{k}\ket{q_\mathbf{k}}\bra{q_\mathbf{k}},
\end{align}
where 
\begin{align}
    \hat{S}_+(l) = \sum_{n=-\infty}^{\infty} (-1)^n \int_{2nl}^{(2n+1)l}\mathrm{d}\mathbf{k}\ket{q_\mathbf{k}}\bra{q_\mathbf{k}+l}
\end{align}
is the spin-step operator and 
\begin{align}
    \hat{S}_-(l) = \hat{S}_+(l)^\dagger.
\end{align}
We are most interested in the two point correlation functions since they are used in the Bell operator. The two points spin operators,$\sxx(l),\szz(l)$, are derived in Appendix A.1 of \cite{Martin:2016tbd}. We will not rewrite the operators here since no real insight is gained due to their non-analytic nature. 

\begin{figure}
\includegraphics[width = 0.9\textwidth]{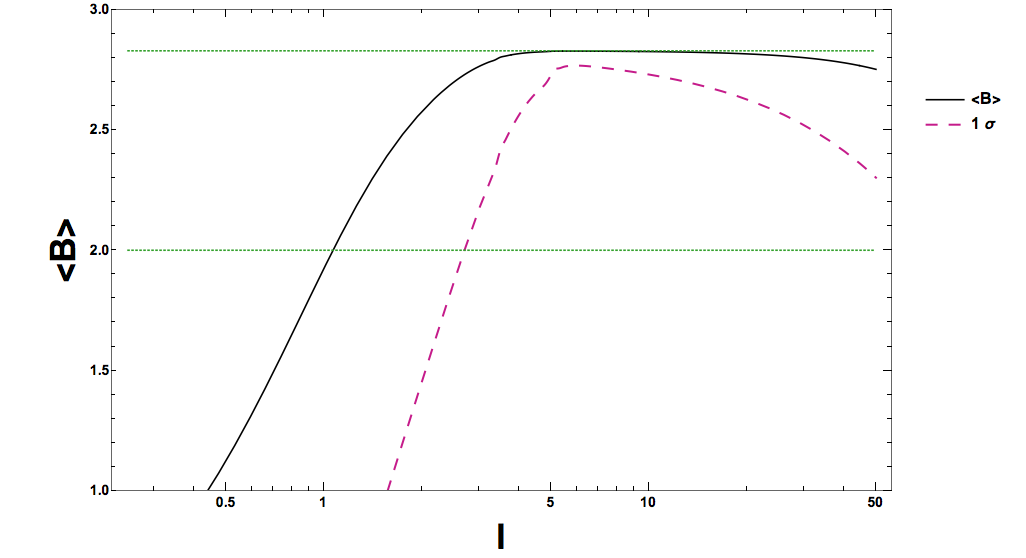}
\caption{Dependence of variance on coarse-graining parameter $l$. For a given choice of $r$ and $\varphi$ there exits a optimal choice of coarse-graining to minimize the variance.}
\label{fig:Larsson_var_r=5}
\end{figure}

 For the optimal angles we look at the dependency of the variance on the choice of coarse-graining parameter for a fixed $r$ and $\varphi$ $r = 5, 2\varphi = - \pi $. We plot this dependence in  Figure \ref{fig:Larsson_var_r=5}. 
From this it is clear that there appears to be an optimal choice of coarse-graining parameter to minimize the variance and maximize the expectation value. This result is consistent with the result shown in Figure 1 of \cite{Martin:2016tbd}. 

\begin{figure}
\includegraphics[width = 0.9\textwidth]{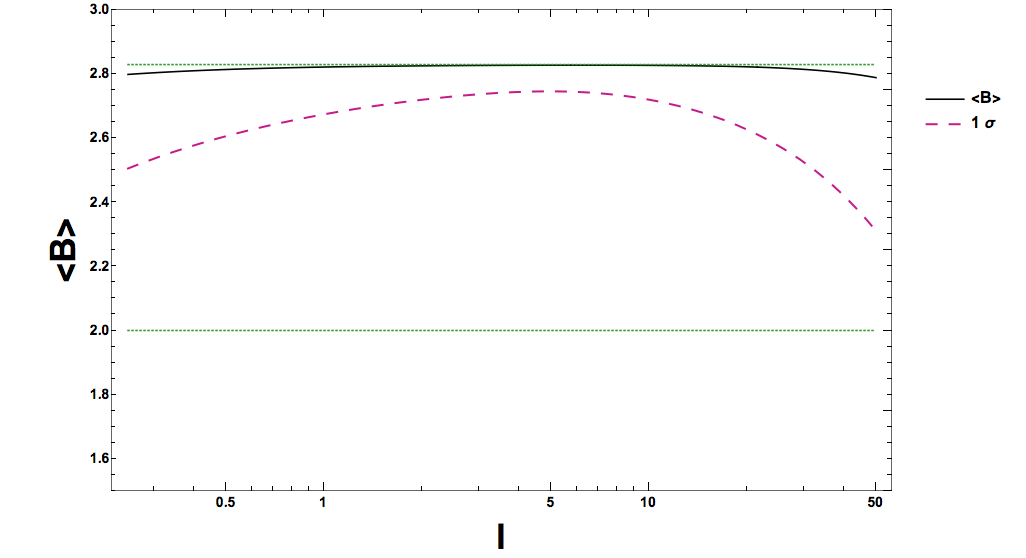}
\caption{Dependence of variance on coarse-graining parameter $l$ in large squeeze limit. When the state becomes highly squeezed, the dependence on coarse graining parameter becomes more relaxed.}
\label{fig:Larsson_var_r_large}
\end{figure}

For cosmological application one is interested in a highly squeezed state $r \sim 50, 2\phi \sim - \pi$.  In the large squeeze limit $ r \gg 1 $ we observe the same behavior as with the other operators, the variance vanishes as the CHSH operator approaches the Cirel'son bound, the highly squeezed limit is plotted in Figure \ref{fig:Larsson_var_r_large}.
These results are still dependent on properly picking the coarse-graining parameter $l$. In the highly squeezed limit if the coarse-graining parameter is not properly chosen the statistical variation will be large and could potential suffer from sample variance. 

It can also be seen that these operators are not as tightly constrained as the previous two sets, and the uncertainty calculation would benefit from using the sample variance. However, as states at the beginning, these operators are of the least interest of the three sets due to the non-analytic form of their two point functions as well as the requirement that $n_s >1$ to realize highly squeezed states. 

\section{Conclusion}
\label{sec:conclusion}

In this paper we extend on the results of \cite{Martin:2016tbd,Martin:2017zxs} by looking at the variance in three sets of pseudo spin operators. For the two analytic sets of operators, the Banaszek-Wodkiewicz and the Gour-Khanna-Mann-Revzen, we find that on highly squeezed states that would  be relevant for observations on the CMB, the variance vanishes and signal observation is potentially strongly seen. For the non-analytic Larsson operators, the variance is a function of both the squeezing parameters as well as the coarse-graining parameter. Therefore, for a state with a certain squeeze the variance bound will depend on the value of the coarse-graining. For the appropriate choice of coarse-graining it is possible to have tight bounds. 

We conclude that the bounds on Bell's violation are tight regardless of sampling, which would serve to further decrease the variance. In some sense as the state becomes highly squeezed the value of CHSH operator becomes a deterministic quantity rather than a stochastic one, since the variation tends to zero as the expectation value saturates the Cirel'son bound. 

While there are other potential limitations to observing Bell's inequality violation \cite{Martin:2017zxs}, statistics is not one of them.

\section*{Acknowledgments}
I thank J\'{e}r\^{o}me Martin and Vincent Vennin for feedback on the initial manuscript as well as Will Kinney for useful discussions.


\bibliographystyle{apsrev4-1}
\bibliography{paper}

\end{document}